\documentclass[12pt]{article}
\usepackage[english]{babel}
\usepackage[utf8]{inputenc}
\usepackage{latexsym}
\usepackage{dsfont}
\usepackage{amsfonts,amsbsy,bm,euscript,mathrsfs}
\usepackage{amssymb,stmaryrd,faktor}
\usepackage[tbtags]{amsmath}
\usepackage{amsthm}
\usepackage[nosort]{cite}
\usepackage{setspace}
\usepackage{graphicx}
\usepackage{color}
\usepackage{mathtools}

\usepackage[toc,page]{appendix}

\usepackage[hidelinks]{hyperref}
\hypersetup{
colorlinks=false,
citecolor= blue,
linkcolor= blue,
urlcolor= blue,
breaklinks=true
}

\allowdisplaybreaks

\numberwithin{equation}{section}

\begin{document}

%%%%%%%%%%%%% TITLE %%%%%%%%%%%%%%%%%%%%%%%%%

\begin{titlepage}
\begin{center}
%\today

\vspace{2.0cm}

{\LARGE  {\fontfamily{lmodern}\selectfont \bf Separation of variables for non-diagonalisable models: \\ \vspace{0.25cm} XXX with twisted boundary conditions}} \\[.2cm]

\vskip 1.5cm

\renewcommand{\thefootnote}{\fnsymbol{footnote}}
\textsc{Juan Miguel Nieto Garc\'ia\footnote{\texttt{juanmiguel.nietogarcia[at]upm.es}}}
\renewcommand{\thefootnote}{\arabic{footnote}}
\setcounter{footnote}{0}

\vskip 0.5cm

\begin{small}
Departamento de Matem\'atica Aplicada a las Tecnolog\'{\i}as de la Informaci\'on \\ y las Comunicaciones, ETSIS de Telecomunicaci\'on, \\ 
Universidad Polit\'ecnica de Madrid \\
C. Nikola Tesla s/n \\
$28031$ Madrid, Spain \\
\end{small}

\end{center}

\vskip 0.7 cm
\begin{abstract}
\vskip0.5cm

\noindent We study the Heisenberg XXX spin chain with twisted periodic boundary conditions with a defective twist matrix. Using quantum separation of variables, we show that the generalised eigenvectors of this system can be characterised in terms of a modified TQ equation that we have named ``Jordanian TQ equation''. We also show that the wavefunction in the separated basis is only separable for true eigenvectors, while the wavefunction for generalised eigenvectors of higher rank is given instead by a sum over weak compositions.

\end{abstract}

\end{titlepage}

%%%%%%%%%%%%%%%%%%% BODY %%%%%%%%%%%%%%%%%%

\section{Introduction}

When looking through the literature for non-diagonalizable or defective integrable models, one finds that nearly all articles consider only the situation where either the Lax matrix, the monodromy matrix, or the transfer matrix becomes non-diagonalizable only at isolated points of the spectral parameter. The case where any of those three matrices is defective for generic values of the spectral parameter has received nearly no attention. This is well reflected in the book on Classical Integrable Systems by O. Babelon, D. Bernard, and M. Talon \cite{Babelon_Bernard_Talon_2003}, where the authors make the following assumption regarding Lax matrices $L(\lambda)$ with poles at the points $\{\lambda_k\}$:
\begin{quote}
``\textbf{Proposition.} Assuming that $L(\lambda)$ has distinct eigenvalues in a neighbourhood of $\lambda_k$, one can perform a regular similarity transformation $g^{(k)}(\lambda)$ diagonalizing $L(\lambda)$ in a vicinity of $\lambda_k$.''
\end{quote}
This proposition implicitly implies that the Lax matrices they consider throughout the book will be diagonalisable for all $\lambda$ except, maybe, at isolated points. This is a widespread assumption, and the techniques used in classical integrability require the Lax matrix to have distinct eigenvalues. In fact, as far as we know, the only example in the literature of a classical monodromy matrix that is non-diagonalisable for generic values of the spectral parameter is the one appearing in the context of the string theory in Newton-Cartan backgrounds \cite{Fontanella:2022wfj, Fontanella:2026gaq}.\footnote{Defective Lax matrices also appear in Ref. \cite{Meessen:2015nla} from rewriting a differential equation as a Lax equation. However, these Lax matrices do not depend on a spectral parameter, and thus lay outside our discussion.}

The situation is a bit better in the case of Quantum Integrable Systems. On the one hand, quantum integrability methods also do not work for non-diagonalisable models. The standard versions of the Bethe Ansatz (either the coordinate, algebraic, modified algebraic, off-diagonal or analytic) implicitly assume diagonalisability, as they build eigenvectors of the Hamiltonian or of the transfer matrix. On the other hand, the need for non-diagonalisable transfer matrices was hinted at already in the late 1980s, when it was shown that quantum groups at roots of unity have indecomposable representations \cite{PASQUIER1990523, ConciniKac, Keller1991, Sun_1991}. Jordan blocks were also known to appear when performing a scaling limit of the XXZ spin chain after applying a non-diagonalisable similarity transformation \cite{Kulish2009}. A better understanding of non-diagonalisable quantum integrable systems came from works on representation theory of Temperley–Lieb algebras and quantum groups at roots of unity \cite{Read:2007qq, Gainutdinov:2013tja}, although these works are more focused on representation theory and the application to Logarithmic Conformal Field Theories than on finding solving techniques using integrability.

In recent years, there has been a growing interest in non-diagonalisable quantum integrable models, especially from the perspective of the AdS/CFT correspondence. There has been work on the eclectic spin chain \cite{Ahn:2020zly, Ahn:2021emp, NietoGarcia:2021kgh, Ahn:2022snr, NietoGarcia:2022kqi}, which arises from the one-loop dilatation operator of the fishnet field theory \cite{Gurdogan:2015csr, Ipsen:2018fmu}; and on Jordanian and Groenewold-Moyal Drinfel'd twists \cite{Borsato:2025smn, Driezen:2025dww, Borsato:2026ypo}, which are dual to some Yang-Baxter deformations in the AdS/CFT correspondence \cite{vanTongeren:2015uha}. Other quantum models with a defective transfer matrix studied in the literature are the XXZ spin chain with open boundary conditions at roots of unity \cite{Gainutdinov:2016pxy} and the class 5 and class 6 deformations of the XXX spin chain \cite{NietoGarcia:2023jeb, deLeeuw:2025sfs}.

Despite such interest, none of these integrable models have been analysed purely using integrability techniques. As regular Bethe Ansatz techniques fail to compute the full basis of generalised eigenvectors, other routes are required. In the articles cited in the previous paragraph we can find many different approaches: combinatorial proofs, symmetry arguments, brute-force triangularisation or treating the non-diagonalisable model as a limit of a diagonalisable model. There are nonetheless three articles that deserve special mention. The first, Ref. \cite{Maillet:2018bim}, analyses the Heisenberg XXX spin chain with twisted boundary conditions using the separation of variables method. In that article, the authors compute the eigenvectors for generic twist, which includes the case where the transfer matrix is non-diagonalisable. However, the authors do not explore the generalised eigenvectors of the model, making the analysis incomplete. The second, Ref. \cite{Driezen:2025izd}, focuses on solving the XXX$_{-1/2}$ model with a non-abelian Jordanian Drinfel'd twist via the TQ equations. This model is known to have Jordan blocks if we consider the polynomial representation, but the authors consider a more general representation where the transfer matrix becomes diagonalisable. Nevertheless, the authors found that the Drinfel'd twist affects the asymptotic behaviour of the Q functions but it does not change the TQ equation. The third, Ref. \cite{Guica:2017mtd}, maps a sector of the dipole deformation of $\mathcal{N}=4$ SYM to an XXX$_{-1/2}$ model with twisted boundary conditions. Similarly to the previous articles we have mentioned, the authors focus on the non-polynomial representations, for which the model is diagonalisable.

In this article we present what we believe is the first construction of generalised eigenvectors of a transfer matrix based solely on integrability techniques. We show that generalised eigenstates can be obtained from a modified TQ equation that we have called \emph{Jordanian TQ equation}. As a test for our proposal, we apply our construction to the specific case of the XXX spin chain with twisted periodic boundary conditions. This spin chain provides a good testing ground because we can construct the eigenvectors using known methods, like the Algebraic Bethe Ansatz or the usual TQ equation, but we cannot do that for the generalised eigenvectors. Combining our Jordanian TQ equation with a co-vector basis obtained from the method of Quantum Separation of Variables, we are able to find the eigenvalues and generalised eigenvectors of this spin chain.

The article is organised as follows. In section 2, we review the Heisenberg XXX spin chain with twisted boundary conditions. In section 3, we show how to modify the method of Separation of Variables for the rational $\mathfrak{gl}(2)$ model to work for a non-diagonalisable transfer matrix. In particular, we find that the usual assumption regarding the separability of the wavefunction in the separated basis only holds for generalised eigenvectors of rank 1, that is, for true eigenvectors. For generalised eigenvectors of higher rank, the wavefunction is given by a sum over weak compositions. In section 4, we apply the construction to the Heisenberg XXX spin chain with a Jordan-block twist matrix. Section 5 closes this paper with a summary of our results and a discussion on future directions of research.

\section{Heisenberg XXX spin chain with twisted boundary conditions} \label{twistdescription}

The Heisenberg spin chain is a quantum-mechanical model originally developed by W. Heisenberg to explain the magnetic properties of solids using a Hamiltonian that makes it energetically favourable for adjacent spins to be aligned. If we consider the case of a circular lattice of $L$ spin $\frac12$ particles, the isotropic or XXX Heisenberg Hamiltonian is given by
\begin{equation}
    H=\frac{1}{4} \sum_{i=1}^L \vec{S}_i \cdot \vec{S}_{i+1} \,, \label{HeisenbergHamiltonian}
\end{equation}
where $\vec{S}_i$ is the spin operator acting on site $i$ of the lattice,\footnote{For spin $\frac12$ particles, the spin operator $\vec{S}_i$ takes the form of Pauli matrices $\vec{\sigma}$ acting on lattice site $i$ and identity in any other site, $\vec{S}_i=\underbrace{\mathbb{I}\otimes \dots \otimes \mathbb{I}}_{i-1\text{ times}}\otimes \frac{\vec{\sigma}}{2} \otimes \mathbb{I}\otimes \dots \otimes \mathbb{I}$.} and the site $L+1$ is periodically identified with site $1$. In our case, we are interested in twisted periodic boundary conditions, so we consider the following identification
\begin{equation}
    \vec{S}_{L+1}=\Theta\vec{S}_1 \,.
\end{equation}
where $\Theta$ is a generic $2\times2$ matrix.

Nowadays, the Heisenberg model is not studied so much as a model of magnetism but as one of the simplest quantum integrable systems. To find the spectrum and eigenstates of this Hamiltonian using integrability, first we define the following R-matrix that acts on two spin $\frac12$ states
\begin{equation}
    R_{a_1,a_2}=u \mathbb{I}_{a_1,a_2}-\eta P_{a_1,a_2} \,, \label{Rmatrix}
\end{equation}
where $\mathbb{I}$ is the identity operator and $P$ is the permutation operator, $P (|x\rangle \otimes |y\rangle)=|y\rangle \otimes |x\rangle$. From the R-matrix, we construct its associated monodromy matrix ${\bf T} (u)$ and transfer matrix $\boldsymbol{\tau} (u)$ as follows
\begin{align}
    {\bf T} (u)&=T(u) \Theta_0=R_{0,1}(u-\theta_1) R_{0,2}(u-\theta_2) \dots R_{0,L-1}(u-\theta_{L-1}) R_{0,L}(u-\theta_L) \Theta_0  \notag \\
    &=\begin{pmatrix}
        A(u) & B(u) \\
        C(u) & D(u)
    \end{pmatrix}_0 \Theta_0  = \begin{pmatrix}
        {\bf A}(u) & {\bf B}(u) \\
        {\bf C}(u) & {\bf D}(u)
    \end{pmatrix}_0  \, , \\
    \boldsymbol{\tau} (u)&=\text{tr}_0 \left[ {\bf T}(u) \right]={\bf A}(u)+{\bf D} (u) \, , \label{transfer}
\end{align}
where the subindices indicate in which space the operators act. Spaces $1$ to $L$ represent the spins in the circular lattice, and thus are isomorphic to $\mathbb{C}^2$, while space $0$ is an auxiliary space that we have chosen to have the same dimensionality as the physical spins. The subindex in the trace $\text{tr}_0$ indicates that it is taken only over the auxiliary space. The variable $u$ is called spectral parameter and the variables $\theta_i$ are called inhomogeneities. To simplify our computations, we will assume that the inhomogeneities fulfil the following condition
\begin{equation}
    \theta_i \neq \theta_j + n \eta \qquad \forall i,j \qquad \forall n\in \mathbb{Z} \,. \label{inhcondition}
\end{equation}
For later convenience, we have distinguished between the monodromy matrix with the twist, ${\bf T}$, and without the twist, $T$.

The R-matrix fulfils the Yang-Baxter equation
\begin{equation}
	R_{a_1, a_2} (u - v ) R_{a_1, a_3} (u -w) R_{a_2,a_3} (v -w) = R_{a_2,a_3} (v -w) R_{a_1, a_3} (u -w) R_{a_1, a_2} (u - v ) \, , \label{YBE}
\end{equation}
while the twist matrix, having no dependence on the spectral parameter, fulfils
\begin{equation}
    [R_{a_1, a_2} (u - v ), \Theta_{a_1} \Theta_{a_2} ]=0\,.
\end{equation}
From these two identities, we can show that both the untwisted monodromy matrix $T$ and the twisted monodromy matrix $\bf T$ fulfil the following equations, called RTT relations
\begin{align}
	R_{a_1, a_2} (u - v ) { T}_{a_1} (u ) { T}_{a_2} (v ) &= { T}_{a_2} (v ) { T}_{a_1} (u ) R_{a_1, a_2} (u - v ) \, , \label{RTTuntwist} \\
    R_{a_1, a_2} (u - v ) {\bf T}_{a_1} (u ) {\bf T}_{a_2} (v ) &= {\bf T}_{a_2} (v ) {\bf T}_{a_1} (u ) R_{a_1, a_2} (u - v ) \, . \label{RTT}
\end{align}
The RTT relations can be used to write an algebra for the operators ${\bf A}$, ${\bf B}$, ${\bf C}$ and ${\bf D}$ that make the entries of the monodromy matrix, and similarly for the untwisted ones. Furthermore, they imply that the transfer matrix commutes with itself for different values of the spectral parameter
\begin{equation}
    \boldsymbol{\tau}  (u) \boldsymbol{\tau} (v)=\boldsymbol{\tau} (v) \boldsymbol{\tau} (u) \,.
\end{equation}
Thus, if we formally expand $\boldsymbol{\tau} (u)$ in Taylor series in $u$, we find a commuting family of operators. Among those, we can find the Hamiltonian \eqref{HeisenbergHamiltonian} if we set $\theta_i=\theta_j$ $\forall i,j$. Thus, diagonalising the transfer matrix corresponds to finding a basis of eigenvectors that diagonalises every member of this family of operators simultaneously and, along the way, the Hamiltonian.

For the case of a diagonal twist,  $\Theta=$ diag $(\delta_1,\delta_2)$, we can apply the usual Algebraic Bethe Ansatz (ABA) to diagonalise the transfer matrix. For pedagogical reviews on the ABA, we refer to \cite{Faddeev:1996iy, Levkovich-Maslyuk:2016kfv, 2018arXiv180407350S} and chapter 7 of \cite{Arutyunov_2026}. In this article, we are only interested in the fact that the eigenstates of the transfer matrix $\boldsymbol{\tau}(u)$ can be constructed from a suitably chosen state, called the pseudo-vacuum and usually taken to be the state in which all spins of the lattice point up, by the repeated action of the ${\bf B}$ operator at specific values
\begin{equation}
    | v_1 , v_2, \dots , v_N \rangle= {\bf B} (v_1) {\bf B} (v_2) \dots {\bf B} (v_N) | 0 \rangle \,. \label{Bethestates}
\end{equation}
These states are called Bethe states and the elements of the set $\{v_i\}$ are called Bethe roots. These Bethe roots are obtained by solving the Bethe Ansatz equations
\begin{equation}
    \prod_{j=1}^L \frac{v_k - \theta_j - \eta}{v_k - \theta_j}= \frac{\delta_2}{\delta_1} \prod_{\substack{l=1\\l\neq k}}^{N} \frac{v_k - v_l - \eta}{v_k - v_l +\eta} \,.
\end{equation}

The case of triangular twist can still be solved using the ABA provided the twist is diagonalisable. The case of generic diagonalisable twist is more involved and cannot be solved using the ABA, as the usual pseudovacua of the ABA (both the state with all spins up and the state with all spins down) cease to be eigenstates of the transfer matrix for most of these twists. There exist nevertheless a Modified Algebraic Bethe Ansatz (MABA) \cite{Belliard:2018pvg} that can be applied in this case.

This leaves only the case of non-diagonalisable twist, which we plan to address in this article. If we consider a twist of Jordan-block form
\begin{equation}
    \Theta=\begin{pmatrix}
        1 & q \\
        0 & 1
    \end{pmatrix} \,, \label{Jtwist}
\end{equation}
by direct computation, we find that the transfer matrix is non-diagonalisable already at $L=2$. Thus, the best we can do is to find a set of vectors such that
\begin{equation}
    \boldsymbol{\tau}(u) |w^{(n)}_j\rangle=\Lambda_j (u) |w^{(n)}_j\rangle+ \sum_{k=1}^{n-1}\Xi_j^{(n,k)} (u) |w^{(k)}_j\rangle \,, \qquad \boldsymbol{\tau}(u) |w^{(1)}_j\rangle=\Lambda (u) |w^{(1)}_j\rangle \,. \label{rankndef}
\end{equation}
We call $|w^{(n)}_j\rangle$ a generalised eigenvector of rank $n$. A set of generalised eigenvectors related in such way is called Jordan chain, indicated by the subindex $j$. From this definition we can see that a generalised eigenvector of rank $n$ satisfies
\begin{equation}
    \left[ \vphantom{\frac12}\boldsymbol{\tau}(u) -\Lambda_j (u) \mathbb{I}\right]^n |w^{(n)}_j\rangle=0 \,, \qquad \left[ \vphantom{\frac12}\boldsymbol{\tau}(u) -\Lambda_j (u) \mathbb{I}\right]^{n-1} |w^{(n)}_j\rangle \neq 0 \,. \label{rankndef2}
\end{equation}
It is important to mention that the functions $\Xi_j^{(n,k)} (u)$ are normally considered non-physical because we can set them to nearly any value by appropriately normalising and linearly combining the generalised eigenvectors.\footnote{Perhaps due to this, we have not found a name for the functions $\Xi_j^{(n,k)} (u)$ in the literature. We have decided to call them \emph{Jordan-chain coefficients}.} The most common choice is to set $\Xi_j^{(n,n-1)} (u)=1$ and $\Xi_j^{(n,k)} (u)=0$, $\forall k<n-1$. This is not the case for us. We want to find a single set of generalised eigenvectors $|w^{(n)}_j\rangle$ that makes $\boldsymbol{\tau}(u)$ upper triangular for all values of $u$. This condition requires that all the generalised eigenvectors have to be independent of $u$. As a consequence, we are only allowed to change their normalisation by a $u$-independent factor, and linear combinations are restricted to have $u$-independent coefficients. Thus, we cannot change the functional dependence of $\Xi_j^{(n,k)} (u)$ with respect to the spectral parameter $u$.

We may wonder if it is always possible to find a $u$-independent basis of generalised eigenvectors. The answer is yes, thanks to the integrability of the problem. Because the transfer matrix commutes with itself for any values of the spectral parameter and it is a polynomial in $u$, its building blocks commute with themselves. As commuting matrices can be simultaneously triangularised over an algebraically closed field, they have a common invariant flag. This flag can be used as the set of generalised eigenvector independent of $u$ that triangularises $\boldsymbol{\tau}(u)$.

For the identity twist, $\Theta=\mathbb{I}$, the Heisenberg XXX spin chain has a global $\mathfrak{gl}(2)$ symmetry that creates a degeneracy between the states. Bethe states \eqref{Bethestates} with finite Bethe roots are highest-weight states with respect to this symmetry. Bethe states with one or more infinite Bethe roots are descendants, in the sense that they can be obtained by applying lowering operators of the $\mathfrak{gl}(2)$ symmetry to highest-weight states. It is not difficult to show that the eigenvectors for the Heisenberg spin chain with the Jordan-block twist \eqref{Jtwist} are in one-to-one correspondence with the highest-weight eigenvectors of the model with identity twist. In fact, because RTT equations do not depend on the twist matrix $\Theta$ and the pseudo-vacuum $ |0 \rangle $ is an eigenstate of $\boldsymbol{\tau} (u)$ with the Jordan-block twist \eqref{Jtwist}, the ABA construction still holds and all the eigenvectors have the form \eqref{Bethestates}. In contrast, the ABA is not capable of constructing the generalised eigenvectors, but there is also a one-to-one correspondence between descendants for the identity twist and generalised eigenvectors for the Jordan-block twist \eqref{Jtwist}. A heuristic explanation is to notice that
\begin{equation}
    \boldsymbol{\tau} (u)={\bf A}(u) + {\bf D}(u)=A(u)+D(u)+q\,C(u) \,, \label{twistedtau}
\end{equation}
for the above Jordan-block twist. Because $A(u)+D(u)$ is the transfer matrix for the identity twist and $C(u)$ is a nilpotent operator that always reduces the number of magnons of a state, states that triangularise $\boldsymbol{\tau}(u)$ can be constructed starting from states that diagonalise the transfer matrix for the identity twist by looking for a linear combination of vectors with fewer magnons that transform the starting state into a generalised eigenvector of the transfer matrix for the Jordan-block twist. The reason why highest-weight eigenvectors always give us eigenvectors and descendants always give us generalised eigenvectors is that, given a descendant state, there is always a state with fewer magnons that shares the same eigenvalue for the identity twist. This is not the case for the highest-weight states provided the inhomogeneities fulfil the condition \eqref{inhcondition}. Once we realise that, the proof follows from the fact that an upper triangular matrix that is not diagonal is diagonalisable if all its diagonal entries are distinct, and it is not diagonalisable if all its diagonal entries are equal. For a practical example on a similar setting, we refer to section 4.2 of \cite{Borsato:2026ypo}. For the twisted Heisenberg model, we will show the explicit computation up to states with two magnons in section \ref{Algebraic}.

\section{Separation of variables for non-diagonalisable transfer matrices} \label{Sov}

One possible method to derive the eigenvectors and the Bethe equations for the XXX spin chain with a generic twist is the quantum separation of variables created by Sklyanin \cite{Sklyanin:1991ss, Sklyanin:1992eu, Sklyanin:1992sm, 10.1143/PTPS.118.35}, and heavily developed in the last decade. Some comprehensive reviews on the modern perspective of the method of separation of variables are \cite{Maillet:2018bim, Ryan:2021duf, Levkovich-Maslyuk:2025ipl} and to chapter 9 of Ref. \cite{Arutyunov_2026}. Although the case of a non-diagonalisable twist was originally studied in \cite{Maillet:2018bim}, the generalised eigenvectors are not studied in that article. In this section, we show that the method of separation of variables can be modified to give us information regarding the generalised eigenvectors. The procedure is essentially the same as for a diagonal twist, except for two specific points: ${\bf B}(u)$ is already a polynomial of degree $L$ in $u$ and the wavefunction is not separable. Therefore, we only present its implementation for the non-diagonalisable twist, highlighting the differences where relevant.

The main idea behind the method of quantum separation of variables for the Heisenberg spin chain is to find a basis where the $\bf B$ operator is diagonal. Because ${\bf B}(u)$ is a polynomial of degree $L$ in the variable $u$~\footnote{In the case of diagonal twist, ${\bf B}(u)\propto B(u)$ is a polynomial of degree $L-1$. This problem is bypassed by performing a similarity transformation of the monodromy matrix. In our case ${\bf B}(u)=B(u)+qA(u)$ is already diagonalisable and of degree $L$ in $u$, so there is no need for a similarity transformation.} and the RTT equations \eqref{RTT} imply that
\begin{equation}
    [{\bf B} (u),{\bf B} (v)]=0 \,,
\end{equation}
we can construct the following family of commuting operators ${\bf x}_k$
\begin{equation}
    {\bf B} (u)={\bf B}_L \prod_{k=1}^L (u - {\bf x}_k) \,,
\end{equation}
where ${\bf B}_L$ is the operator in ${\bf B} (u)$ accompanying the power $u^L$, that is
\begin{equation}
    {\bf B}_L=\lim_{u\to\infty} u^{-L} \,\textbf{B}(u) \,. \label{BL}
\end{equation}
For values of the inhomogeneities fulfilling the condition~\eqref{inhcondition}, the individual operators ${\bf x}_k$ are not simple but their combined spectrum is simple, that is, the set of eigenvalues $\{{ x}_k\}_{k=1}^L$ with respect to the set of operators $\{{\bf x}_k\}_{k=1}^L$ is different for different eigenvectors.

The operators conjugate to ${\bf x}_k$, which we will denote by ${\bf z}^\pm_j$, are constructed from the ${\bf A} (u)$ and ${\bf D} (u)$ operators. As ${\bf A} (u)$ and ${\bf D} (u)$ are polynomials of degree $L$ in $u$, we can write them as follows
\begin{equation}
    {\bf A} (u)=\sum_{k=0}^L u^k {\bf A}_k \,,\qquad {\bf D} (u)=\sum_{k=0}^L u^k {\bf D}_k \,.
\end{equation}
Then, the operators ${\bf z}^\pm_j$ are defined as
\begin{equation}
    {\bf z}^+_j=\sum_{k=0}^L {\bf x}_j^k {\bf A}_k \,,\qquad {\bf z}^-_j=\sum_{k=0}^L {\bf x}_j^k {\bf D}_k \,,
\end{equation}
notice that ordering matters here because these operators do not commute. Conversely, using again that ${\bf A} (u)$ and ${\bf D} (u)$ are polynomials of degree $L$ in $u$, we can use the Lagrange interpolation formula to write them in terms of the ${\bf z}^\pm_j$ operators
\begin{align}
    {\bf A} (u)={\bf A}_L \prod_{j=1}^L (u - {\bf x}_j) + \sum_{j=1}^L \prod_{k\neq j}\frac{u-{\bf x}_k}{{\bf x}_j-{\bf x}_k}{\bf z}^+_j \,, \label{Az}\\
    {\bf D} (u)={\bf D}_L \prod_{j=1}^L (u - {\bf x}_j) + \sum_{j=1}^L \prod_{k\neq j}\frac{u-{\bf x}_k}{{\bf x}_j-{\bf x}_k}{\bf z}^-_j \,, \label{Dz}
\end{align}
where ${\bf A}_L$ and ${\bf D}_L$ are defined similarly to ${\bf B}_L$ in equation \eqref{BL}, but with respect to the appropriate operator.

From the RTT equations \eqref{RTT}, we obtain a set of commutation relations between these operators
\begin{align}
    &[{\bf x}_i ,{\bf x}_j]=[{\bf z}^\pm_i,{\bf z}^\pm_j]=0\,, \quad \forall i,j \notag \\
    &[{\bf z}^\pm_i,{\bf z}^\mp_j]=0 \,,\quad \forall i\neq j \label{xzcomm}\\
    &{\bf z}^\pm_i {\bf x}_j =({\bf x}_j\pm \eta \,\mathbb{I} \delta_{ij})\,{\bf z}^\pm_i \,, \quad \forall i,j \notag
\end{align}
while from the definition of the quantum determinant $\Delta(u)={\bf A} (u){\bf D} (u+\eta)-{\bf B} (u){\bf C} (u+\eta)$ we obtain that
\begin{equation}
    {\bf z}^+_j{\bf z}^-_j=\Delta ({\bf x}_j) \,, \qquad {\bf z}^-_j{\bf z}^+_j=\Delta({\bf x}_j-\eta \,\mathbb{I}) \,. \label{zzcomm}
\end{equation}

The next step is to construct a basis in the dual Hilbert space that simultaneously diagonalises all ${\bf x}_k$ operators
\begin{equation}
    \langle x_1 , \dots , x_L| {\bf x}_k=\langle x_1 , \dots , x_L| x_k \qquad \forall k \,. \label{xonvectors}
\end{equation}
From the commutation relations \eqref{xzcomm}, it is clear that ${\bf z}^\pm_j$ behave like ladder operators in this basis
\begin{equation}
    \langle x_1 , \dots ,x_j , \dots , x_L| {\bf z}^\pm_j=\langle x_1 , \dots ,x_j \pm \eta, \dots , x_L| \Delta^\pm_j(\vec{x})  \,, \label{zonvectors}
\end{equation}
where $\Delta^\pm_j(\vec{x})$ are functions of the eigenvalues, which we collectively have denoted as $\vec{x}$. If we denote the shift of the eigenvalue $x_j$ by $\eta$ as $\vec{x}+\eta\,\vec{e}_j$, equation~\eqref{zzcomm} implies that these functions satisfy
\begin{equation}
    \Delta^+_j (\vec{x}) \Delta^-_j(\vec{x}+\eta\,\vec{e}_j)=\Delta(x_j) \,, \qquad \Delta^-_j (\vec{x}) \Delta^+_j(\vec{x}+\eta\,\vec{e}_j)=\Delta(x_j-\eta) \,. 
\end{equation}
For consistency, we also require $\Delta^\pm_j(\vec{x})$ to vanish if $x_j\pm\eta$ is not among the eigenvalues of ${\bf x}_j$. These restrictions are not enough to completely fix these functions. This is expected, as they can be arbitrarily modified by changing the normalisation of our states.\footnote{Our discussion below equation~\eqref{rankndef2} does not apply to these states.} A particularly convenient choice is
\begin{equation}
    \Delta^\pm_j (\vec{x})=\Delta^\pm(x_j) \,, \qquad \text{where} \qquad   \Delta^+ (u)=\Delta^-(u-\eta)=\prod_{k=1}^L (u-\theta_k-\eta) \,.\label{Deltachoice}
\end{equation}

We can now compute the wavefunction of a generalised eigenstate of the transfer matrix
\begin{equation}
    \psi^{(n)}_j(\vec{x})=\langle x_1 , \dots , x_L | w^{(n)}_j \rangle=\langle \vec{x} | w^{(n)}_j \rangle \,.
\end{equation}
This can be done by computing the matrix element $\langle \vec{x} | \boldsymbol{\tau}(x_i) | w^{(n)}_j\rangle$ in two different ways. On the one hand, $| w^{(n)}_j\rangle$ is a generalised eigenstate of the transfer matrix, so it fulfils \eqref{rankndef}. On the other hand, we know how the transfer matrix $\boldsymbol{\tau}(u)$ acts on $\langle \vec{x} |$ by combining \eqref{Az}, \eqref{Dz}, \eqref{xonvectors} and \eqref{zonvectors}. From that, we get the following equation for the wavefunction
\begin{equation}
    \Lambda_j(x_i) \psi^{(n)}_j(\vec{x}) + \sum_{k=1}^{n-1}\Xi_j^{(n,k)} (x_i) \psi^{(k)}_j(\vec{x})=\Delta_i^-(\vec{x}) \psi^{(n)}_j(\vec{x}-\eta \,\vec{e}_i)+\Delta_i^+(\vec{x}) \psi^{(n)}_j(\vec{x}+\eta \,\vec{e}_i) \,. \label{unseparatedeq}
\end{equation}
Although this equation looks like the usual wavefunction equation for separated variables (see, e.g. equation (9.278) in \cite{Arutyunov_2026}) with an additional Jordanian term, solving it beyond the eigenvectors is a bit more involved.

For the case of eigenvectors (or generalised eigenvectors of rank 1), where we have no Jordan-chain coefficients, this set of equations is solved by assuming that the wavefunction is factorised\footnote{As the eigenstates of the transfer matrix are still Bethe states of the form \eqref{Bethestates}, it is natural that the wavefunction separates in the basis that diagonalises the $\bf B$ operator.}
\begin{equation}
    \psi_j^{(1)} (\vec{x})=\prod_{i=1}^L \phi_{i,j}^{(1)} (x_i) \,,  \label{ansatzQ1}
\end{equation}
giving us $L$ separated equations for each $\phi_{i,j}^{(1)} (x_i)$, all with the same structure. Because all these equations have the same structure, we can solve them simultaneously by considering instead the following TQ equation
\begin{equation}
    \boldsymbol{\Lambda}(u) Q^{(1)}(u) =\Delta^-(u) Q^{(1)}(u-\eta)+\Delta^+(u) Q^{(1)}(u+\eta) \,, \label{homTQ}
\end{equation}
where $\Delta^\pm(u)$ are given by \eqref{Deltachoice}, and set $\phi_{i,j}^{(1)} (x_i)=Q_j^{(1)}(x_i)$.

For the generalised eigenvector of rank 2, the presence of the Jordanian term forbids us from applying the same logic if we assume separability of the wavefunction. Instead, to rewrite the wavefunction equation as a TQ equation, we have to consider the following ansatz
\begin{equation}
    \psi_j^{(2)} (\vec{x})=\sum_{k=1}^L \left[ \phi_{k,j}^{(2)} (x_k)\prod_{\substack{i=1\\i\neq k}}^L \phi_{i,j}^{(1)} (x_i) \right] \,. \label{ansatzQ2}
\end{equation}
We can now take the wavefunction equation \eqref{unseparatedeq}, substitute this ansatz, and divide the equation by $\psi_j^{(1)} (\vec{x})$, giving us
\begin{align}
    \Lambda_j(x_i) \sum_{k=1}^L \frac{\phi^{(2)}_{k,j} (x_k)}{\phi^{(1)}_{k,j} (x_k)}+\Xi_j^{(2,1)}&=\Delta_i^-(\vec{x}) \left[ \frac{\phi^{(2)}_{i,j} (x_i-\eta)}{\phi^{(1)}_{i,j} (x_i)} + \sum_{l\neq i} \frac{\phi^{(2)}_{l,j} (x_l)}{\phi^{(1)}_{l,j} (x_l)}\,\frac{\phi^{(1)}_{i,j} (x_i-\eta)}{\phi^{(1)}_{i,j} (x_i)} \right] \\
    &+ \Delta_i^+(\vec{x}) \left[ \frac{\phi^{(2)}_{i,j} (x_i+\eta)}{\phi^{(1)}_{i,j} (x_i)} + \sum_{l\neq i} \frac{\phi^{(2)}_{l,j} (x_l)}{\phi^{(1)}_{l,j} (x_l)}\,\frac{\phi^{(1)}_{i,j} (x_i+\eta)}{\phi^{(1)}_{i,j} (x_i)} \right] \notag \\
    &=\Delta_i^-(\vec{x}) \frac{\phi^{(2)}_{i,j} (x_i-\eta)}{\phi^{(1)}_{i,j} (x_i)}+\Delta_i^+(\vec{x}) \frac{\phi^{(2)}_{i,j} (x_i+\eta)}{\phi^{(1)}_{i,j} (x_i)}+\Lambda_j (x_i) \sum_{l\neq i} \frac{\phi^{(2)}_{l,j} (x_l)}{\phi^{(1)}_{l,j} (x_l)} \,, \notag
\end{align}
where, to get the last line, we have used the fact that $\phi_{i,j}^{(1)}(x_i)$ solves the homogeneous equation \eqref{homTQ}. Rearranging the terms, all the dependence on $\phi^{(2)}_{k,j}(x_k)$ with $k\neq i$ disappears and it becomes an equation only for $\phi^{(2)}_{i,j}(x_i)$. As with eigenvectors, we can now set $\phi_{i,j}^{(2)}(x_i)=Q_j^{(2)}(x_i)$, where $Q_j^{(2)}(u)$ fulfils the following inhomogeneous TQ equation
\begin{equation}
    \boldsymbol{\Lambda}(u) Q^{(2)}(u) + \boldsymbol{\Xi}^{(2,1)} (u) Q^{(1)}(u)=\Delta^-(u) Q^{(2)}(u-\eta)+\Delta^+(u) Q^{(2)}(u+\eta) \,, \label{inhomTQ2}
\end{equation}
with the same $\boldsymbol{\Lambda}(u)$ and $Q^{(1)}$ that solve the homogeneous equation \eqref{homTQ}. Due to its form, we call this equation \emph{Jordanian TQ equation}.

For the generalised eigenvector of rank 3, let us consider the ansatz
\begin{equation}
    \psi_j^{(3)} (\vec{x})=\sum_{k=1}^L \left[ \phi_{k,j}^{(3)} (x_k)\prod_{\substack{i=1\\i\neq k}}^L \phi_{i,j}^{(1)} (x_i) \right]+\kappa\sum_{\substack{k,l=1\\k< l}}^L \left[ \phi_{k,j}^{(2)} (x_k)\phi_{l,j}^{(2)} (x_l)\prod_{\substack{i=1\\i\neq k,l}}^L \phi_{i,j}^{(1)} (x_i) \right] \,, \label{ansatzQ3}
\end{equation}
where $\kappa$ is a constant, and repeat the same steps as for the case of the generalised eigenvector of rank 2
\begin{align}
    &\Lambda_j(x_i) \sum_{k=1}^L \frac{\phi^{(3)}_{k,j} (x_k)}{\phi^{(1)}_{k,j} (x_k)}+\kappa\, \Lambda_j(x_i) \sum_{\substack{k,l=1\\k< l}}^L  \frac{\phi_{k,j}^{(2)} (x_k)\phi_{l,j}^{(2)} (x_l)}{\phi_{k,j}^{(1)} (x_k)\phi_{l,j}^{(1)} (x_l)} +\Xi_j^{(3,2)} \sum_{k=1}^L \frac{\phi^{(2)}_{k,j} (x_k)}{\phi^{(1)}_{k,j} (x_k)} +\Xi_j^{(3,1)}  \label{computationpsi3}\\
    &=\Delta_i^-(\vec{x}) \left[ \frac{\phi^{(3)}_{i,j} (x_i-\eta)}{\phi^{(1)}_{i,j} (x_i)} + \sum_{l\neq i} \frac{\phi^{(3)}_{l,j} (x_l)}{\phi^{(1)}_{l,j} (x_l)}\,\frac{\phi^{(1)}_{i,j} (x_i-\eta)}{\phi^{(1)}_{i,j} (x_i)} + \kappa \sum_{l\neq i}  \frac{\phi_{i,j}^{(2)} (x_i-\eta)\phi_{l,j}^{(2)} (x_l)}{\phi_{i,j}^{(1)} (x_i)\phi_{l,j}^{(1)} (x_l)}  +\right. \notag \\
    &+\left.\kappa\sum_{\substack{k,l=1\\i\neq k< l\neq i}}^L  \frac{\phi_{k,j}^{(2)} (x_k)\phi_{l,j}^{(2)} (x_l)}{\phi_{k,j}^{(1)} (x_k)\phi_{l,j}^{(1)} (x_l)} \,\frac{\phi^{(1)}_{i,j} (x_i-\eta)}{\phi^{(1)}_{i,j} (x_i)}\right] + \Delta_i^+(\vec{x}) \left[ \frac{\phi^{(3)}_{i,j} (x_i+\eta)}{\phi^{(1)}_{i,j} (x_i)} + \sum_{l\neq i} \frac{\phi^{(3)}_{l,j} (x_l)}{\phi^{(1)}_{l,j} (x_l)}\,\frac{\phi^{(1)}_{i,j} (x_i+\eta)}{\phi^{(1)}_{i,j} (x_i)}  \right. \notag \\
    &+\left.  \kappa\sum_{l\neq i}  \frac{\phi_{i,j}^{(2)} (x_i+\eta)\phi_{l,j}^{(2)} (x_l)}{\phi_{i,j}^{(1)} (x_i)\phi_{l,j}^{(1)} (x_l)}+ \kappa\sum_{\substack{k,l=1\\i\neq k< l\neq i}}^L  \frac{\phi_{k,j}^{(2)} (x_k)\phi_{l,j}^{(2)} (x_l)}{\phi_{k,j}^{(1)} (x_k)\phi_{l,j}^{(1)} (x_l)} \,\frac{\phi^{(1)}_{i,j} (x_i+\eta)}{\phi^{(1)}_{i,j} (x_i)}\right] \notag \\
    &=\Delta_i^-(\vec{x}) \frac{\phi^{(3)}_{i,j} (x_i-\eta)}{\phi^{(1)}_{i,j} (x_i)}+\Delta_i^+(\vec{x}) \frac{\phi^{(3)}_{i,j} (x_i+\eta)}{\phi^{(1)}_{i,j} (x_i)}+\Lambda_j (x_i) \sum_{l\neq i} \frac{\phi^{(3)}_{l,j} (x_l)}{\phi^{(1)}_{l,j} (x_l)}  \notag \\
    &+\kappa \, \Lambda_j (x_i) \sum_{l\neq i}  \frac{\phi_{i,j}^{(2)} (x_i)\phi_{l,j}^{(2)} (x_l)}{\phi_{i,j}^{(1)} (x_i)\phi_{l,j}^{(1)} (x_l)} + \kappa \, \Xi_j^{(2,1)} \sum_{l\neq i}  \frac{\phi_{l,j}^{(2)} (x_l)}{\phi_{l,j}^{(1)} (x_l)} + \kappa \, \Lambda_j (x_i)  \sum_{\substack{k,l=1\\i\neq k< l\neq i}}^L  \frac{\phi_{k,j}^{(2)} (x_k)\phi_{l,j}^{(2)} (x_l)}{\phi_{k,j}^{(1)} (x_k)\phi_{l,j}^{(1)} (x_l)} \,. \notag
\end{align}
Now, provided we have $\Xi_j^{(3,2)}(u)=\kappa \,\Xi_j^{(2,1)} (u)$, we can rearrange the terms to make it an equation only for $\phi^{(3)}_{i,j}(x_i)$. Mutatis mutandis, we can set $\phi_{i,j}^{(3)} (x_i)=Q_j^{(3)}(x_i)$, where $Q_j^{(3)}(u)$ fulfils the following inhomogeneous TQ equation
\begin{multline}
    \boldsymbol{\Lambda}(u) Q^{(3)}(u)+ \boldsymbol{\Xi}^{(3,2)} (u) Q^{(2)}(u) + \boldsymbol{\Xi}^{(3,1)} (u) Q^{(1)}(u)=\\=\Delta^-(u) Q^{(3)}(u-\eta)+\Delta^+(u) Q^{(3)}(u+\eta) \,, \label{inhomTQ3}
\end{multline}
with the same $\boldsymbol{\Lambda}(u)$, $Q^{(1)}$ and $Q^{(2)}$ that solve equations \eqref{homTQ} and \eqref{inhomTQ2}. This again can be considered a Jordanian TQ equation.

The wavefunction equations for generalised eigenvectors of higher rank can be reduced to Jordanian TQ equations iteratively following a similar procedure. The appropriate ansatz for the general case would be 
\begin{equation}
    \psi_j^{(n)} (\vec{x})=\sum_{\{\lambda\}} \kappa_{\{\lambda\}} \prod_{k=1}^L \phi_{k,j}^{(1+\lambda_k)} (x_k) \,,
\end{equation}
where the sum over $\{\lambda\}$ represents a sum over all the possible weak compositions of $n-1$ into $L$ parts.\footnote{Weak compositions of $n-1$ into $L$ parts are defined as sets of $L$ non-negative integers $\{\rho_i\}_{i=1}^L$ such that $\sum_{i=1}^L \rho_i=n-1$, where two sets that differ in the order of their terms are considered different. For example, there are six different weak compositions of $2$ into $3$ parts, given the sets $\{1,1,0\}$, $\{1,0,1\}$, $\{0,1,1\}$, $\{2,0,0\}$, $\{0,2,0\}$ and $\{0,0,2\}$. Weak partitions are defined similarly to weak compositions, but sets that differ in the order of their terms are considered the same. For the previous example, there are only two weak partitions, $\{1,1,0\}$ and $\{2,0,0\}$.} The constants $\kappa_{\{\lambda\}}$ depend only on the partition associated with the composition we are considering. This means that the coefficients $\kappa_{\{\lambda\}}$ are the same for any two compositions that are related by a permutation of their entries. For simplicity, we can set $\kappa=1$ for the partition with the largest amount of zeros. Then, provided the Jordan-chain coefficients fulfil
\begin{equation}
    \Xi_j^{(n,k)}=\kappa_{\{\lambda\}}\Xi_j^{(n-1,k-1)} \qquad \forall k>1 \,,
\end{equation}
for the appropriate $\kappa_{\{\lambda\}}$, we can set $\phi_{i,j}^{(n)} (x_i) =Q_j^{(n)}(x_i)$, where $Q_j^{(n)}(u)$ fulfils the following Jordanian TQ equation
\begin{equation}
    \boldsymbol{\Lambda}(u) Q^{(n)}(u) + \sum_{k=1}^{n-1} \boldsymbol{\Xi}^{(n,k)} (u) Q^{(k)}(u)=\Delta^-(u) Q^{(n)}(u-\eta)+\Delta^+(u) Q^{(n)}(u+\eta) \,, \label{inhomTQn}
\end{equation}
with the same $\boldsymbol{\Lambda}(u)$ that solves the homogeneous equation \eqref{homTQ}.

Although one might be tempted to compare the Jordanian term of \eqref{inhomTQn} with the inhomogeneous terms of the inhomogeneous TQ equations appearing in the off-diagonal Bethe Ansatz \cite{Cao:2013qxa, Cao:2013bca} and in the MABA \cite{Belliard:2018pvg}, the two have entirely different origins and interpretations. The inhomogeneous TQ equations arise in models with broken $U(1)$ symmetry, which prevents us from finding a reference state to build the standard ABA construction. If we act with the transfer matrix into the usual ABA-type reference states, we get additional terms due to this broken $U(1)$. If we perform an ABA-like construction of the eigenvectors, these terms can be incorporated into the eigenvalues and Bethe equations. Upon rewriting the resulting Bethe equations as a TQ equation, the terms associated with the breaking of the $U(1)$ symmetry appear as an inhomogeneous contribution. In comparison, the Jordanian term in \eqref{inhomTQn} arises explicitly from the non-diagonalisability of the transfer matrix, hence the dependence on a lower-rank Q function. It is an intrinsic property of the transfer matrix and not a consequence of using a specific basis to diagonalise it.

We should emphasize that at no point of the derivation have we specified that we are considering the twisted XXX spin chain. We only have used the diagonalisability of the $\bf B$ operator\footnote{In principle, this condition could be relaxed to demanding that the transfer matrix to be basis generating if we follow instead the construction from Ref. \cite{Maillet:2018bim}.} and the commutation relations associated to the R-matrix \eqref{Rmatrix}. Therefore, we propose that the generalised eigenvectors of any non-diagonalisable transfer matrix constructed from the rational $\mathfrak{gl}(2)$ model where we can construct separated variables can be described using a Jordanian TQ equation.

\section{TQ equations for the Jordanianly twisted XXX spin chain}

In this section we explicitly show how the construction from section~\ref{Sov} works for the XXX spin chain with the Jordan-block twist \eqref{Jtwist}. We will first perform the construction of separated variables in detail for $L=2$, comparing the generalised eigenvectors, eigenvalues and Jorcan-chain coefficients we get from brute-force triangularisation with the ones we obtained using the Jordanian TQ equation. After that, we compute the generalised eigenvectors with up to two magnons for arbitrary length by computing commutation relations using the Algebraic Bethe Ansatz and compare these results with the ones we obtained using the Jordanian TQ equation. To distinguish the different results, we will write $\boldsymbol{\Lambda}$ and $\boldsymbol{\Xi}$ for the eigenvalues and Jordan-chain coefficients computed using the TQ equations and $\Lambda$ and $\Xi$ for the eigenvalues and Jordan-chain coefficients computed using other methods.

\subsection{$L=2$}

\subsubsection{Brute-force triangularisation}

For length $L=2$ and twist \eqref{Jtwist}, the monodromy matrix ${\bf T}(u)$ has the following form
{\footnotesize 
\makeatletter
\renewcommand{\maketag@@@}[1]{\hbox{\m@th\normalsize\normalfont #1}}
\makeatother
\begin{align}
    {\bf A} (u) &= \begin{pmatrix}
        (u-\theta_1-\eta)(u-\theta_2-\eta) & 0 & 0 & 0 \\
        0 & (u-\theta_1-\eta)(u-\theta_2) & 0 & 0 \\
        0 & \eta^2 & (u-\theta_1)(u-\theta_2-\eta) & 0 \\
        0 & 0 & 0 & (u-\theta_1)(u-\theta_2) 
    \end{pmatrix} \,, \notag \\
    {\bf B} (u) &=\begin{pmatrix}
        q(u-\theta_1-\eta)(u-\theta_2-\eta) & 0 & 0 & 0 \\
        -\eta(u-\theta_1-\eta) & q(u-\theta_1-\eta)(u-\theta_2) & 0 & 0\\
-\eta(u-\theta_2) & q\eta^2 & q(u-\theta_1)(u-\theta_2-\eta) & 0 \\
        0 & -\eta(u-\theta_2-\eta) & -\eta(u-\theta_1) & q(u-\theta_1)(u-\theta_2) 
    \end{pmatrix} \,, \notag \\
    {\bf C} (u) &=\begin{pmatrix}
        0 & -\eta(u-\theta_1) & -\eta(u-\theta_2-\eta) & 0 \\
        0 & 0 & 0 & -\eta(u-\theta_2)\\
        0 & 0 & 0 & -\eta(u-\theta_1-\eta)\\
        0 & 0 & 0 & 0
    \end{pmatrix} \,, \\
    {\bf D} (u) &=\begin{pmatrix}
        (u-\theta_1)(u-\theta_2) & -q\eta(u-\theta_1) & -q\eta(u-\theta_2-\eta) & 0 \\
        0 & (u-\theta_1)(u-\theta_2-\eta) & \eta^2 & -q\eta(u-\theta_2) \\
        0 & 0 & (u-\theta_1-\eta)(u-\theta_2) & -q\eta(u-\theta_1-\eta) \\
        0 & 0 & 0 & (u-\theta_1-\eta)(u-\theta_2-\eta)
    \end{pmatrix}     \,. \notag
\end{align}}
On the one hand, we can compute the dual eigenvectors that diagonalise the ${\bf B} (u)$ operator,
\begin{align}
    &\langle \theta_1 , \theta_2|=(1,-q,-q,q^2) \,, & &\langle \theta_1+\eta , \theta_2|=(1,-q,0,0)\,, \notag \\
    &\langle \theta_1 , \theta_2+\eta|=\left(1,-\frac{q\,\eta}{\theta_1 - \theta_2 + \eta},-\frac{q(\theta_1-\theta_2)}{\theta_1 - \theta_2 + \eta},0 \right) \,, & &\langle \theta_1+\eta , \theta_2+\eta|=(1,0,0,0) \,. 
\end{align}
These vectors indeed satisfy \eqref{zonvectors} with the choice \eqref{Deltachoice}.

On the other hand, the generalised eigenvectors of the transfer matrix are
\begin{align}
    &|w^{(1)}_1\rangle=\left( q, -\frac{2\eta}{\theta_1 - \theta_2 -\eta} , \frac{2\eta}{\theta_1 - \theta_2 -\eta},0 \right) \,, & \Lambda_1(u) &=\prod_i(u-\theta_i) + \prod_i(u-\theta_i-\eta) - 2\eta^2 \,, \notag \\
    &|w^{(1)}_2\rangle=\left( 1, 0, 0, 0 \right) \,, & \Lambda_2(u) &=\prod_i(u-\theta_i) + \prod_i(u-\theta_i-\eta)  \,, \notag \\
    &|w^{(2)}_2\rangle=\left( \alpha_0, \alpha_1, \alpha_1, 0 \right) \,, & \Xi_2^{(2,1)} &=-\alpha_1 q \eta (2u-\eta -\theta_1 - \theta_2)\,, \\
    &|w^{(3)}_2\rangle=\left( \beta_0, \beta_1 -\beta_2 q(\theta_1 -\theta_2 +\eta), \beta_1, 2\eta \beta_2 \right) \,, & \Xi_2^{(3,2)}&=-\frac{\beta_2}{\alpha_1} q \eta^2 (2u-\eta -\theta_1 - \theta_2) \,.  \notag\\
    &\mathrlap{ \Xi_2^{(3,1)}=q^2 \eta \beta_2 (\theta_1 -\theta_2 + \eta) (u-\theta_1)+q\eta \frac{\alpha_0 \beta_2 \eta - \alpha_1 \beta_1}{\alpha_1} (2u-\eta -\theta_1 - \theta_2)  \,,}  \notag 
\end{align}
where $\alpha_1$ and $\beta_2$ are two arbitrary non-vanishing constants and $\alpha_0$, $\beta_1$ and $\beta_0$ are arbitrary constants. These coefficients parameterise the freedom we have to normalise the generalised eigenvectors and to linearly combine them with generalised eigenvectors of lower rank with the same eigenvalue, respectively. We should also point out that, as we discussed below equation~\eqref{rankndef}, we cannot set $\Xi_2^{(3,1)}=0$ and $\Xi_2^{(2,1)}=\Xi_2^{(3,2)}=1$ while keeping the generalised eigenvectors independent of $u$. Thus, the Jordan-chain coefficients are physically meaningful in this setting.

The scalar product between these two sets of vectors is collected in the following matrix, where the rows are the different $\langle x_1, x_2|$ vectors and the columns are the different $|w_j\rangle$ vectors
\begin{equation}
    \langle x_1 , x_2 | w_j\rangle=\begin{pmatrix}
        q & 1 & \alpha_0-2q\alpha_1 & \beta_0 - 2q\beta_1 +q^2(\theta_1 - \theta_2 +3\eta)\beta_2 \\
        q \frac{\theta_1 - \theta_2 +\eta}{\theta_1 - \theta_2 - \eta}  & 1 & \alpha_0-q\alpha_1 & \beta_0 - q\beta_1 +q^2(\theta_1 - \theta_2 +\eta)\beta_2\\
        q \frac{\theta_1 - \theta_2 - \eta}{\theta_1 - \theta_2 +\eta} & 1 & \alpha_0-q\alpha_1 & \beta_0 - q\beta_1 +q^2 \eta \beta_2\\
        q & 1 & \alpha_0 & \beta_0
    \end{pmatrix} \,. \label{wavefunctionL2}
\end{equation}

\subsubsection{Jordanian TQ equation}

Let us now compute the eigenvalues, the Jordan-chain coefficients and the generalised eigenvectors using the Jordanian TQ equations, and compare them with the results obtained above using brute-force triangularisation. We start by computing the eigenvectors of the transfer matrix by looking for solutions of the homogeneous TQ equation. For length $L=2$, the solutions to \eqref{homTQ} are
\begin{align}
    Q^{(1)}_1(u) &=u-\frac{\theta_1 + \theta_2 +\eta}{2} \,, & \boldsymbol{\Lambda}_1(u) &=\prod_{i=1}^2 (u-\theta_i) + \prod_{i=1}^2 (u-\theta_i-\eta) - 2\eta^2 \,,\\
    Q^{(1)}_2(u)&=1 \,, & \boldsymbol{\Lambda}_2(u) &=\prod_{i=1}^2 (u-\theta_i) + \prod_{i=1}^2 (u-\theta_i-\eta) \,.
\end{align}
We can immediately check that the eigenvalues match the ones computed in the previous sections. Using the ansatz \eqref{ansatzQ1} to compute the wavefunctions from the Q function, we can check that they perfectly match the generalised eigenvectors of rank 1 we found up to a normalisation factor.

Let us now move to Jordanian TQ equations associated with $Q^{(1)}_2(u)=1$. To compute the Q function associated with the generalised eigenvector of rank 2, we have to substitute the appropriate values for $Q^{(1)}_2(u)$ and $\boldsymbol{\Lambda}_2(u)$ into \eqref{inhomTQ2}, and solve for $Q^{(2)}_2(u)$ and $\boldsymbol{\Xi}^{(2,1)}_2(u)$. We find
\begin{align}
    Q^{(2)}_2(u)&=a_1 u+a_0 \,, & &\boldsymbol{\Xi}^{(2,1)}_2(u) =-a_1 \eta^2 (2u-\eta -\theta_1 - \theta_2)  \,,
\end{align}
where $a_1$ and $a_0$  are coefficients that are not fixed by the Jordanian TQ equation. This is expected. We cannot fix $a_0$ because the linear combination of a generalised eigenvector of rank $2$ and an eigenvector with the same eigenvalue is also a generalised eigenvector of rank $2$. We also cannot fix $a_1$ because multiplying a generalised eigenvector by a scalar gives us the same generalised eigenvector, and only modifies the Jordan-chain coefficients.

We clearly observe a matching between the Jordan-chain coefficients $\boldsymbol{\Xi}^{(2,1)}_2$ and $\Xi^{(2,1)}_2$ up to normalisation coefficients. In particular
\begin{equation}
    \boldsymbol{\Xi}^{(2,1)}_2(u)=\frac{a_1\eta}{\alpha_1 q} \Xi^{(2,1)}_2(u) \,. \label{matchingxi2}
\end{equation}
As we discussed above, the functions $\Xi$ carry physical information in our setting. Thus, the matching between them is a concrete test for our construction.

Let us now compute the wavefunction associated with this Q function. Substituting $Q^{(1)}_2(u)$ and $Q^{(2)}_2(u)$ into the ansatz \eqref{ansatzQ2}, we get
\begin{equation}
    \psi^{(2)}_2(x_1,x_2)= Q^{(2)}_2 (x_1) Q^{(1)}_2 (x_2)+Q^{(1)}_2 (x_1) Q^{(2)}_2 (x_2)= a_1 (x_1+x_2)+2a_0 \,.
\end{equation}
We can check that this wavefunction fulfils the following two properties
\begin{align}
    &\psi^{(2)}_2(\theta_1+\eta,\theta_2)=\psi^{(2)}_2(\theta_1,\theta_2+\eta) \,, \\ 
    &\psi^{(2)}_2(\theta_1,\theta_2)-\psi^{(2)}_2(\theta_1+\eta,\theta_2)=\psi^{(2)}_2(\theta_1+\eta,\theta_2)-\psi^{(2)}_2(\theta_1+\eta,\theta_2+\eta) \,,
\end{align}
which are also fulfilled by the wavefunction \eqref{wavefunctionL2}. As both have two undetermined coefficients and fulfil these two constraints, they are the same up to matching the normalisation factors. In fact, if we set
\begin{equation}
    \alpha_0=a_1 (\theta_1+\theta_2+2\eta)+2a_0 \,, \qquad q\alpha_1=a_1\eta \,, \label{mapQ2}
\end{equation}
the matching between the two wavefunctions is perfect. Notice that, using the second of these conditions into \eqref{matchingxi2}, we have $\boldsymbol{\Xi}^{(2,1)}_2(u)= \Xi^{(2,1)}_2(u)$.

Let us move now to the generalised eigenvector of rank 3. Substituting into \eqref{inhomTQ3} the appropriate values for $Q^{(1)}_2(u)$, $Q^{(2)}_2(u)$ and $\boldsymbol{\Lambda}_2(u)$, and solving for $Q^{(3)}_2(u)$ and $\boldsymbol{\Xi}^{(3,1)}_2(u)$, we obtain
\begin{align}
    Q^{(3)}_2(u)&=b_2u^2+b_1u+b_0 \,, \\
    \boldsymbol{\Xi}^{(3,1)}_2(u) + (a_1 u+a_0) \boldsymbol{\Xi}^{(3,2)}_2(u) &=b_2 \eta^2[\theta_1 \theta_2 +(\theta_1 + \eta)(\theta_2+\eta)-2u^2]-b_1 \eta^2(2u-\eta -\theta_1 - \theta_2)  \notag \,. 
\end{align}
Similarly to the generalised eigenvector of rank 2, the $b_i$ are coefficients that are not fixed by the Jordanian TQ equations. This is again expected for similar reasons.

First, as we discussed below \eqref{computationpsi3}, we need $\boldsymbol{\Xi}^{(3,2)}_2(u)=\kappa \, \boldsymbol{\Xi}^{(2,1)}_2(u)$ to get the Jordanian TQ equation \eqref{inhomTQ3} from the wavefunction equation \eqref{unseparatedeq}. Substituting this condition and demanding that $\boldsymbol{\Xi}^{(3,1)}_2(u)$ be linear in $u$,\footnote{The off-diagonal entries of the transfer matrix $\boldsymbol{\tau}(u)$ are all linear in $u$, so it is sensible to assume that the Jordan-chain functions $\boldsymbol{\Xi}^{(n,k)}$ are, at most, linear.} we get the condition $b_2=\kappa a_1^2$.

The wavefunction for the generalised eigenvector of rank 3 is given by the ansatz \eqref{ansatzQ3}, which in this case takes the form
\begin{align}
    \psi^{(3)}_2(x_1,x_2)&= Q^{(3)}_2 (x_1) Q^{(1)}_2 (x_2)+Q^{(1)}_2 (x_1) Q^{(3)}_2 (x_2)+ \kappa \, Q^{(2)}_2 (x_1) Q^{(2)}_2 (x_2) \,, \label{wavefunctionQ3} \\
    &= b_2 (x_1^2+x_2^2)+\kappa a_1^2 x_1 x_2+(\kappa a_0 a_1 + b_1) (x_1+x_2)+\kappa a_0^2+2b_0 \,. \notag
\end{align}
We can check that the map
\begin{align}
b_0 &=\frac{ a_1^2\left[ \eta \beta_0 - q \beta_1 (\theta_1 + \theta_2 +2\eta) +q^2 \beta_2 (\theta_1 + \eta)(\theta_1 + 2\theta_2 + 3 \eta) \right] -q^2 a_0^2 \beta_2 }{2\eta a_1^2} \,,
\\
b_1 &=\frac{ q a_1 \beta_1 -q^2 a_0 \beta_2 -q^2 a_1 \beta_2(2\theta_1 + \theta_2 + 3 \eta) }{\eta a_1} \,,
\\
b_2 &=\frac{q^2\beta_2}{\eta}\,, \qquad \kappa =\frac{q^2\beta_2}{a_1^2\eta}\,,
\end{align}
together with the map \eqref{mapQ2}, gives us a perfect matching between the wavefunctions \eqref{wavefunctionQ3} and \eqref{wavefunctionL2} and between $\boldsymbol{\Xi}^{(3,1)}_2(u)$ and ${\Xi}^{(3,1)}_2(u)$.

Thus, we have shown that we can construct the eigenvalues, the Jordan-chain coefficients and the generalised eigenvectors from the generalised Q functions computed via the Jordanian TQ equation. Furthermore, they match perfectly with the ones obtained from brute-force triangularisation once the correct identification of the free coefficients is done. This provides a successful test of our construction.

\subsection{One and two magnons at any $L$}

\subsubsection{Algebraic triangularisation} \label{Algebraic}

From the RTT equation~\eqref{RTTuntwist} we can extract the following commutation relations for the operators of the monodromy matrix with no twist
\begin{align}
    &A(u) B(v)=\frac{u-v+\eta}{u-v} B(v) A(u) -\frac{\eta}{u-v} B(u) A(v) \,, \notag \\
    &D(u) B(v)=\frac{u-v-\eta}{u-v} B(v) D(u) +\frac{\eta}{u-v} B(u) D(v) \,, \label{commrelABA} \\
    &[C(u),B(v)]=\frac{\eta}{u-v} \left[ \vphantom{\frac12} A(u) D(v)-A(v) D(u) \right] \,. \notag
\end{align}
We also choose a pseudo-vacuum $| 0 \rangle$ that has the following properties
\begin{align}
    &A (u)| 0 \rangle= \Delta^+(u) | 0 \rangle \,, & &D (u)| 0 \rangle= \Delta^-(u) | 0 \rangle \,, \\
   & C(u)| 0 \rangle=0 \,, & &B (u)| 0 \rangle\neq \Delta^-(u) | 0 \rangle \,, \notag
\end{align}
where $\Delta^\pm$ are defined in \eqref{Deltachoice}. Using these relations, we can show that $| 0 \rangle$ is an eigenstate of the twisted transfer matrix $\boldsymbol{\tau} (u)$ for the Jordan-block twist \eqref{Jtwist}
\begin{equation}
    \boldsymbol{\tau} (u) |0\rangle=\left[ \Delta^+(u)+\Delta^-(u) \right] |0\rangle=\left[ \prod_{i=1}^L \left( u-\theta_i - \eta \right) + \prod_{i=1}^L \left( u-\theta_i \right)  \right] |0\rangle \,,
\end{equation}
while for the state obtained by acting once with a $B(u)$ operator on the pseudo-vacuum we get
\begin{align}
    [A(u)+D(u)+q\,C(u)] B(v) |0\rangle&= \left( \frac{u-v+\eta}{u-v} \Delta^+(u)+ \frac{u-v-\eta}{u-v} \Delta^-(u) \right) B(v) |0\rangle \notag \\
    &- \frac{\eta}{u-v}\left[ \Delta^+(v)-\Delta^-(v) \right] B(u) |0\rangle \notag \\
    &+\frac{q\,\eta}{u-v}\left[ \Delta^+(u) \Delta^-(v)-\Delta^+(v) \Delta^-(u) \right]  |0\rangle \,. \label{comrelonemag}
\end{align}
If we find a finite $\tilde{v}$ such that $\Delta^+(\tilde{v})=\Delta^-(\tilde{v})$, then we can construct the following eigenvector
\begin{equation}
    \boldsymbol{\tau} (u)\left[ B(\tilde{v})+ q \Delta^+(\tilde{v}) \right] |0\rangle= \left( \frac{u-\tilde{v}+\eta}{u-\tilde{v}} \Delta^+(u)+ \frac{u-\tilde{v}-\eta}{u-\tilde{v}} \Delta^-(u) \right) \left[ B(\tilde{v})+ q \Delta^+(\tilde{v}) \right] |0\rangle \,.
\end{equation}
Because $\Delta^+(\tilde{v})=\Delta^-(\tilde{v})$ is a polynomial equation of degree $L-1$, this gives us $L-1$ different eigenvectors. The attentive reader might wonder why are using the operator $B$ when we discussed that eigenstates can be constructed via the usual ABA using the operator $\bf B$. We can even see that $\left[ B(\tilde{v})+ q \Delta^+(\tilde{v}) \right] |0\rangle = {\bf B}(\tilde{v})|0\rangle $. We follow this approach because it is easier to construct the generalised eigenvectors using the untwisted operators than using the twisted operators, where taking one of the rapidities to infinity gives no new vectors. In fact, if we multiply the equation \eqref{comrelonemag} by a factor $v^{1-L}$ and take the limit $v\to \infty$, we find that the state we obtain is the generalised eigenvector of rank 2 associated to the pseudo-vacuum
\begin{align}
    &\left( \vphantom{\frac12} \boldsymbol{\tau} (u) - [\Delta^+(u)+\Delta^-(u)] \right) \,\, \lim_{v\to \infty} v^{1-L}B(v) |0\rangle =-q\,\eta\left[ \Delta^+(u) - \Delta^-(u) \right]  |0\rangle \,, \label{descendantL}\\
    &\left( \vphantom{\frac12} \boldsymbol{\tau} (u) - [\Delta^+(u)+\Delta^-(u)] \right)^2 \lim_{v\to \infty} v^{1-L}B(v) |0\rangle = 0 \,.
\end{align}

Let us move now to the state created with two $B$ operators. An eigenstate of $\boldsymbol{\tau} (u)$ can be written as the following linear combination
\begin{equation}
    B(v) B(w) |0\rangle + \alpha_1 B(v) |0\rangle + \alpha_2 B(w) |0\rangle +\alpha_3 |0\rangle \,.
\end{equation}
After acting with the transfer matrix $\boldsymbol{\tau} (u)$ on this linear combination, using the commutation relations \eqref{commrelABA}, and demanding that the resulting quantity be proportional to the original vector, we find the following relations
\begin{align}
    \alpha_1&=-q\frac{(u-v-\eta) (v-w-\eta) \Delta^+(w) \Delta^-(u) + (u-v+\eta) (v-w+\eta) \Delta^-(w) \Delta^+(u)}{(v-w) [ (u-v+\eta) \Delta^+ (u) - (u-v-\eta) \Delta^-(u)]} \, , \notag \\
    \alpha_2&=-q\frac{(u-w-\eta) (v-w+\eta) \Delta^+(v) \Delta^-(u) + (u-w+\eta) (v-w-\eta) \Delta^-(v) \Delta^+(u)}{(v-w) [ (u-w+\eta) \Delta^+ (u) - (u-w-\eta) \Delta^-(u)]} \, , \notag\\
    \alpha_3&=\frac{q \,\alpha_1 (u-w) [\Delta^+(u) \Delta^-(v) - \Delta^-(u) \Delta^+(v)]}{(2u-v-w+\eta) \Delta^+(u) - (2u-v-w-\eta)\Delta^-(u) } \notag\\
    &+\frac{q \,\alpha_2 (u-v) [\Delta^+(u) \Delta^-(w) - \Delta^-(u) \Delta^+(w)]}{(2u-v-w+\eta) \Delta^+(u) - (2u-v-w-\eta)\Delta^-(u) }\,, \notag \\
    &\frac{\Delta^+ (v)}{\Delta^-(v)}=\frac{\Delta^-(w)}{\Delta^+ (w)}=\frac{v-w-\eta}{v-w+\eta}\,.
\end{align}
The last of the equations is the usual two-magnon Bethe equation. If we impose these Bethe equations, the coefficients $\alpha_i$ heavily simplify
\begin{align}
    &\alpha_1=q\frac{\tilde{v}-\tilde{w}+\eta}{\tilde{v}-\tilde{w}} \Delta^-(\tilde{w})=q\frac{\tilde{v}-\tilde{w}-\eta}{\tilde{v}-\tilde{w}} \Delta^+(\tilde{w}) \, , \\
    &\alpha_2=q\frac{\tilde{v}-\tilde{w}-\eta}{\tilde{v}-\tilde{w}} \Delta^-(\tilde{v})=q\frac{\tilde{v}-\tilde{w}+\eta}{\tilde{v}-\tilde{w}} \Delta^+(\tilde{v}) \, , \\
    &\alpha_3=q^2 \Delta^-(\tilde{v}) \Delta^-(\tilde{w})=q^2 \Delta^+(\tilde{v}) \Delta^+(\tilde{w}) \,, 
\end{align}
We also find that the eigenvalue associated with this state is
\begin{equation}
    \Lambda(u)=\frac{u-\tilde{v}+\eta}{u-\tilde{v}} \frac{u-\tilde{w}+\eta}{u-\tilde{w}} \Delta^+(u)+ \frac{u-\tilde{v}-\eta}{u-\tilde{v}} \frac{u-\tilde{w}-\eta}{u-\tilde{w}} \Delta^-(u) \,,
\end{equation}
where the hats indicate that the rapidities fulfil the above Bethe equations.

However, for the descendant obtained by taking one of the parameters to infinity, one can check that the na\"ive ansatz
\begin{equation}
    \lim_{w\to \infty} w^{1-L} B(\tilde{v}) B(w) |0\rangle +\beta_1 B(\tilde{v}) |0\rangle + \beta_2 \lim_{w\to \infty} w^{1-L} B(w) |0\rangle + \beta_3 |0\rangle  \,,
\end{equation}
where $\Delta^+(\tilde{v})=\Delta^-(\tilde{v})$, cannot be made into an eigenstate of $\boldsymbol{\tau} (u)$. This happens because the commutation relations create an unwanted term proportional to $B(u) |0\rangle$ that cannot be eliminated.\footnote{Substituting $B(\tilde{v})$ for $B(x)$ for arbitrary $x$ solves this problem, but the resulting vector cannot be made into an eigenvector of $\boldsymbol{\tau} (u)$ unless $x=\tilde{v}$, which takes us back to the original problem.} We can nevertheless show that
\begin{align}
    &\boldsymbol{\tau} (u) \left[ \lim_{w\to \infty} w^{1-L} B(\tilde{v}) B(w) |0\rangle + \dots  \right] \notag \\
    &= \left( \frac{u-\tilde{v}+\eta}{u-\tilde{v}} \Delta^+(u)+ \frac{u-\tilde{v}-\eta}{u-\tilde{v}} \Delta^-(u) \right) \left[ \lim_{w\to \infty} w^{1-L} B(\tilde{v}) B(w) |0\rangle + \dots \right] \notag \\
    &- q\eta \left( \frac{u-\tilde{v}+\eta}{u-\tilde{v}} \Delta^+(u) - \frac{u-\tilde{v}-\eta}{u-\tilde{v}} \Delta^-(u) \right) B(\tilde{v})  |0\rangle \,,
\end{align}
provided $\Delta^+(\tilde{v})=\Delta^-(\tilde{v})$. Here the ellipses represent a term that only involves vectors from the sector with one magnon. We can see that this cannot be the full solution because the parenthesis in the last line, which should correspond with ${\Xi}^{(2,1)} (u)$, is not a polynomial in $u$. Nevertheless, we will see that this result is not far from the actual one.

The situation is the same for the second descendant of the vacuum. The na\"ive ansatz
\begin{equation}
    \lim_{v\to \infty}\lim_{w\to \infty} (vw)^{1-L} B(v) B(w) |0\rangle +\gamma_1 \lim_{w\to \infty} w^{1-L} B(w) |0\rangle + \gamma_3 |0\rangle  \,,
\end{equation}
will always have an unwanted term $2q\eta^2 B(u) |0\rangle$ that cannot be eliminated. Nevertheless, we can show that
\begin{align}
    &\boldsymbol{\tau} (u) \left[ \lim_{v\to \infty}\lim_{w\to \infty} (vw)^{1-L} B(v) B(w) |0\rangle + \dots  \right] \notag \\
    &= \left[  \Delta^+(u)+  \Delta^-(u) \right] \left[ \lim_{v\to \infty}\lim_{w\to \infty} (vw)^{1-L} B(v) B(w) |0\rangle + \dots  \right] \notag \\
    &- 2 q\eta \left(  \Delta^+(u) -  \Delta^-(u) \right)  \left[ \lim_{v\to \infty}\lim_{w\to \infty} w^{1-L}  B(w) |0\rangle + \dots  \right] + \dots \,,
\end{align}
Here the ellipses represent a term that only involves the vacuum and vectors from the sector with one magnon. From this computation we can check that ${\Xi}_2^{(3,2)} (u)=2{\Xi}_2^{(2,1)} (u)$, but not much more.

\subsubsection{Jordanian TQ equation}

Let us now compute the solutions to the Jordanian TQ equations for constant, linear and quadratic Q functions and compare them with the ones obtained above.

The homogeneous TQ equation \eqref{homTQ} has the following solutions
\begin{align}
    Q^{(1)}_1 &= 1 \,,  & \boldsymbol{\Lambda}(u)&=\Delta^-(u)+\Delta^+(u) \,, \\
    Q^{(1)}_2 &= u-\tilde{v} \,,  & \boldsymbol{\Lambda}(u)&=\frac{u-\tilde{v}+\eta}{u-\tilde{v}} \Delta^+(u)+ \frac{u-\tilde{v}-\eta}{u-\tilde{v}} \Delta^-(u) \,, \\
    Q^{(1)}_3 &= (u-\tilde{u})(u-\tilde{w}) \,,  & \boldsymbol{\Lambda}(u)&=\frac{u-\tilde{u}+\eta}{u-\tilde{u}} \frac{u-\tilde{w}+\eta}{u-\tilde{w}} \Delta^+(u)+ \frac{u-\tilde{u}-\eta}{u-\tilde{u}}\frac{u-\tilde{w}-\eta}{u-\tilde{w}} \Delta^-(u) \,. 
\end{align}
The values $\tilde{u}$, $\tilde{v}$ and $\tilde{w}$ are not fixed by the TQ equation. Instead, they are fixed from demanding that $\boldsymbol{\Lambda}(u)$ be a polynomial, giving us an additional constraint equivalent to the Bethe equations. This result matches the computations above.

Let us move now to the Jordanian TQ equation for generalised eigenvectors of rank 2, given by equation \eqref{inhomTQ2}. For the generalised eigenvector associated with the vacuum, we have to solve
\begin{equation}
    [\Delta^-(u)+\Delta^+(u)] Q^{(2)}_1(u) + \boldsymbol{\Xi}_1^{(2,1)} (u)=\Delta^-(u) Q^{(2)}_1(u-\eta)+\Delta^+(u) Q^{(2)}_1(u+\eta) \,.
\end{equation}
If we input the ansatz $Q^{(2)}_1(u)=a_1 u + a_0$, we find that neither $a_1$ nor $a_0$ are fixed by the TQ equation, but $\boldsymbol{\Xi}_1^{(2,1)}$ is fixed to be
\begin{equation}
    \boldsymbol{\Xi}_1^{(2,1)}=a_1\eta [ \Delta^+(u) - \Delta^-(u)] \,.
\end{equation}
This matches \eqref{descendantL} up to a factor of $-q$, which can be extracted from the normalisation $a_1$.

For the generalised eigenvector associated with the one-magnon state, we have to solve
\begin{multline}
    \left( \frac{u-\tilde{v}+\eta}{u-\tilde{v}} \Delta^+(u)+ \frac{u-\tilde{v}-\eta}{u-\tilde{v}} \Delta^-(u) \right) Q^{(2)}_2(u) + \boldsymbol{\Xi}_2^{(2,1)} (u) (u-\tilde{v})=\\
    =\Delta^-(u) Q^{(2)}_2(u-\eta)+\Delta^+(u) Q^{(2)}_2(u+\eta) \,.
\end{multline}
Analysing the asymptotic behaviour of the Jordanian TQ equation, we are allowed to assume that $Q^{(2)}_2(u)$ is a quadratic polynomial in $u$ and that $\boldsymbol{\Xi}_2^{(2,1)} (u)$ is a polynomial of degree $L-1$ in $u$. The equation is easier to solve if we change to the variable $y=u-\tilde{v}$. First, we use the equation at $y=0$ to fix the constant coefficient of $Q^{(2)}_2(u)$, and we use this result to fix $\boldsymbol{\Xi}_2^{(2,1)} (u)$. The final result is
\begin{align}
    Q^{(2)}_2(u) &= b_2 (u-\tilde{v})^2 + b_1(u-\tilde{v}) + b_2 \frac{\Delta^+ (\tilde{v})+\Delta^- (\tilde{v})}{[\Delta^+ (\tilde{v})]'-[\Delta^- (\tilde{v})]'} \,,\\
    \boldsymbol{\Xi}_2^{(2,1)} (u) &=b_2 \eta \left[  \frac{u-\tilde{v}+\eta}{u-\tilde{v}} \Delta^+(u) - \frac{u-\tilde{v}-\eta}{u-\tilde{v}} \Delta^-(u) \right. \notag \\
    &\left.-b_2\eta\frac{\Delta^+ (u)-\Delta^- (u)}{(u-\tilde{v})^2} \,\frac{\Delta^+ (\tilde{v})+\Delta^- (\tilde{v})}{[\Delta^+ (\tilde{v})]'-[\Delta^- (\tilde{v})]'} \right] \,,
\end{align}
where primes indicate derivatives with respect to the argument. We can see that the first two terms of $\boldsymbol{\Xi}_2^{(2,1)} (u)$ match what we found using brute-force with the Algebraic Bethe Ansatz. We can also check that the last term cancels the non-vanishing residue at $u=\tilde{v}$ from the first two terms, making $\boldsymbol{\Xi}_2^{(2,1)} (u)$ a polynomial.

\section{Conclusions}

In this article we have proposed the Jordanian TQ equation, a modification of the TQ equation valid for integrable models with non-diagonalisable transfer matrices. As far as we know, this is the first integrability-based method to be capable of tackling non-diagonalisable transfer matrices. As a test for this novel proposal, we have applied the Jordanian TQ equation to the XXX spin chain with a twist of Jordan-block form. We have shown that we can reproduce the generalised eigenvectors, Jordan-chain coefficients and eigenvalues for the case of length 2, and we can reconstruct the eigenvalues and the Jordan-chain coefficients for any value of the length and two or fewer magnons. We have also shown that the wavefunctions for generalised eigenvectors of rank $n>1$ are not separable, taking instead the form of sums over weak compositions of $n-1$ into $L$ non-negative integers.

The derivation of the Jordanian TQ equation we have presented assumes that we can construct separated variables and we have a rational $\mathfrak{gl}(2)$ R-matrix. Therefore, the most immediate question would be how much we can generalised it. One possible direction would be to go beyond $\mathfrak{gl}(2)$ symmetry. The construction we presented should be immediately generalisable to rational models with higher rank. Therefore, it would be interesting to see if it can be applied to the eclectic spin chain of Refs. \cite{Ipsen:2018fmu, Ahn:2020zly}, which is obtained from the limit of a deformation of a spin chain with $\mathfrak{gl}(3)$ symmetry. Another direction would be to generalise it to trigonometric models, like the XXZ spin chain with open boundary conditions at roots of unity studied in Ref. \cite{Gainutdinov:2016pxy}. A final possibility would be to stay in the rational $\mathfrak{gl}(2)$ and study infinite-dimensional representations in the non-compact cases. The XXX$_{-1/2}$ version of the model we have studied in this article has already been studied in Ref. \cite{Guica:2017mtd}. However, the authors focused on the non-polinomial repersentations, so the non-diagonalisable sector of the theory has not been explored yet.

Drinfel'd twists of the rational $\mathfrak{gl}(2)$ model are also an important source of models we can study. One example would be the XXX$_{-1/2}$ model with a non-abelian Jordanian Drinfel'd twist studied in Refs. \cite{Borsato:2025smn, Driezen:2025dww, Driezen:2025izd}. It was shown in Ref. \cite{Driezen:2025izd} that, for the representation where the coproduct of the spin raising operator is diagonalisable, the TQ equation is unchanged with respect to the untwisted model, yet the asymptotic behaviour of the Q functions is non-trivially modified. It would be interesting to analyse the XXX$_{-1/2}$ model in the polynomial representation, where the transfer matrix is non-diagonalisable. In this case, we expect that we have to also modify the asymptotic behaviour of the Q functions for this representation.

Another Drinfel'd twist of the XXX$_{1/2}$ spin chain is the Class 5 model. This model was originally found in Ref. \cite{DeLeeuw:2019gxe} as part of the classification of all regular difference-form $4\times4$ R-matrices that satisfy the Yang-Baxter equation \eqref{YBE}. Similarly to the model studied in this article, the eigenvectors of the Class 5 model that are in one-to-one correspondence with the highest-weight eigenvectors of the XXX$_{1/2}$ model, while all the generalised eigenvectors are in correspondence with descendants \cite{NietoGarcia:2023jeb}. Sadly, in this case it can be shown that no linear combination of the $A$, $B$, $C$ and $D$ operators is diagonalisable. Thus, the construction of this article has no reason to work. This should not discourage us. Instead of performing separation of variables \emph{à la Sklyanin}, we can construct the dual basis using the method from Ref. \cite{Maillet:2018bim}. This method does not require either of these operators to be diagonalisable. Instead, it requires the transfer matrix to be \emph{basis-generating}, meaning that the transfer matrix is capable of generating a covector basis when acting on a generic dual vector. As this is a weaker condition, it might hold for this model.

There are also technical questions we have not explored in this article. First, we have not explored how general the condition $\Xi_j^{(n,k)}=\kappa_{\{\lambda\}}\Xi_j^{(n-1,k-1)}$ needed to construct the Jordanian TQ equation is. We only know that it seems to hold for the model we are interested in. Another technical question we have not explored is the existence of a second set of Q functions. The TQ equation for the periodic XXX$_{1/2}$ spin chain has two different Q functions tied to each eigenstate, one Q function associated with creating a state from the pseudo-vacuum with all spins up, and a second Q function associated with creating a state from the pseudo-vacuum with all spins down. Here, we have computed one type of Q functions, but we have not studied what happens to the other one. It would also be interesting to see if these two Q functions satisfy a Wronskian Bethe equation.

%%%%%%%%%%%%%%%%%%%%%%%%%%%%%%%%%%%%%%%%%%%%%%%%%%%%%%%%%%%%%%%%%%
%%%%%%%%%%%%%%%%%%%%%%%%%%%%%%%%%%%%%%%%%%%%%%%%%%%%%%%%%%%%%%%%%%

\vspace{8mm}

\centerline{\bf Acknowledgments}

\vspace{2mm}

\noindent
We want to thank Marius de Leeuw, Miguel García Fernández and Rafael Hernández for useful discussion on the topic and comments on the manuscript.

%%%%%%%%%%%%%%%%%%%%%%%%%%%%%%%%%%%%%%%%%%%%%%%%%%%%%%%%%%%%%%%%%%
%%%%%%%%%%%%%%%%%%%%%%%%%%%%%%%%%%%%%%%%%%%%%%%%%%%%%%%%%%%%%%%%%%

%%%%%%%%%%%%%%%% BIBLIOGRAPHY %%%%%%%%%%%%%%%%

\bibliographystyle{nb}

\bibliography{Biblio.bib}

\end{document}